\documentclass[runningheads,a4paper]{llncs}

\usepackage[numbers,sort&compress,sectionbib]{natbib}

\usepackage{amsmath}
\usepackage{amssymb}
\usepackage{graphicx}
\usepackage{epstopdf}

\usepackage{url}
\newcommand{\keywords}[1]{\par\addvspace\baselineskip
\noindent\keywordname\enspace\ignorespaces#1}

\begin{document}

\mainmatter  

\title{Predictability of social interactions}

\author{Kevin S.~Xu%
\thanks{Current affiliation: 3M Corporate Research Laboratory, 
St.~Paul, MN, USA}%
}

\institute{Department of Electrical Engineering and Computer Science,\\
University of Michigan, Ann Arbor, MI, USA\\
\url{xukevin@umich.edu}}

\maketitle

\begin{abstract}
The ability to predict social interactions between people has profound 
applications including targeted marketing and prediction of information 
diffusion and disease propagation. 
Previous work has shown that the location of an individual at any given 
time is highly predictable. 
This study examines the \emph{predictability of social interactions} between 
people to determine whether interaction patterns are similarly predictable. 
I find that the \emph{locations} and \emph{times} of interactions for an 
individual are highly predictable; however, the \emph{other person} the 
individual interacts with is less predictable. 
Furthermore, I show that knowledge of the locations and times of interactions 
has \emph{almost no effect} on the predictability of the other person. 
Finally I demonstrate that a simple Markov chain model is able to achieve 
\emph{close to the upper bound} in terms of predicting the next person 
with whom a given individual will interact. 
\end{abstract}

\section{Introduction}
One of the most important questions in the emerging field of human 
dynamics concerns \emph{predictability}: to what extent is human behavior 
predictable, and how does predictability vary across the population?
Recent technological advances have led to the development of wearable human 
sensors, which are capable of continuously collecting data on an individual's 
movement, activities, and interactions among other features. 
These sensors could allow us to make predictions about many aspects of human 
behavior including social interactions. 
The ability to make such predictions has profound applications such as 
targeted marketing using an individual's social network to predicting how 
diseases transmitted through human contact propagate over time.

This study is motivated by previous work on the predictability of 
human mobility. 
Using cell phone data from $50,000$ individuals, \citet{Song2010} studied the 
\emph{predictability of individuals' locations} using the closest cellular 
tower each time an individual used his phone. 
To capture the predictability of an individual's location over time, the 
authors estimated the \emph{entropy rate} of the time series of his locations 
and found that the real uncertainty in an individual's location at 
any given time is \emph{fewer than two locations}!
Furthermore, there was found to be surprisingly little variability in the 
estimated entropy rate among the population, suggesting that individuals' 
locations over time are, in general, \emph{highly predictable}.

The main question behind this study is as follows: \emph{to what extent are 
individuals' social interactions predictable?} 
I investigate the extent to which three aspects of an individual's 
interactions are predictable: the \emph{physical location} and 
\emph{time} of an interaction and the \emph{other person} 
with whom the individual interacts. 
Similar to \citet{Song2010}, I estimate the \emph{entropy rate} to capture 
predictability. 
In addition, I use a Markov chain model for social interaction 
to evaluate actual prediction performance on two real data sets.

\section{Methodology}

\subsection{Entropy rates}
To capture the predictability of a time series, I utilize the entropy rate. 
First we have the notion of \emph{entropy}, which measures the amount of 
uncertainty in a random variable. 
The entropy of a single random variable $X$ is defined as 
$H(X) = -\sum_{i} p(x_i) \log_2 p(x_i)$, 
where the summation is over all possible outcomes $\{x_i\}$, and $p(x_i)$ 
denotes the probability of outcome $x_i$. 
For two random variables $(X,Y)$, we have the notions of \emph{joint entropy} 
$H(X,Y)$ and \emph{conditional entropy} $H(X|Y)$. 
The joint entropy measures the uncertainty associated with both random 
variables, while 
the conditional entropy measures the uncertainly in one random variable given 
that the value of the other random variable has been observed. 
The joint and conditional entropies are related through the equation 
$H(X|Y) = H(X,Y) - H(Y)$.

The \emph{entropy rate} was first introduced by \citet{Shannon1948} and 
generalizes the notion of entropy to sequences of \emph{dependent} random 
variables. 
For a stationary stochastic process $X = \{X_i\}$, the entropy rate is 
defined as
\begin{equation}
	\label{eq:Def_entropy_rate}
	H(X) = \lim_{n \rightarrow \infty} \frac{1}{n} H(X_1, X_2, \dots, X_n) 
		= \lim_{n \rightarrow \infty} H(X_n | X_1, X_2, \dots, X_{n-1})
\end{equation}
where the first equality holds for all stochastic processes, but the second 
requires stationarity of the process. 
The quantity on the right side of \eqref{eq:Def_entropy_rate} leads to the 
interpretation of entropy rate as the uncertainty in a quantity at 
time $n$ having observed the complete history. 
The entropy rate denotes the average per-variable entropy of each random 
variable in the stochastic process. 
Joint and conditional entropy rates can similarly be defined. 
In this study, I use the entropy rate to characterize the 
\emph{average uncertainty of a quantity at any given time}.

\subsection{Lempel-Ziv complexities}
To calculate the entropy of a random variable $X$, one needs to know the 
probability of each possible outcome $p(x_i)$. 
When these probabilities are not known, one can \emph{estimate} the entropy 
by replacing the probabilities with relative 
frequencies from observed data. 
Estimating the entropy rate of a stochastic process is more involved because 
the random variables are, in general, dependent on one another. 
A suitable estimator of the entropy rate for general stationary stochastic 
processes is the \emph{Lempel-Ziv complexity}. 
Similar to \citet{Song2010}, I use the following Lempel-Ziv complexity to
estimate the entropy rate of a time series:
\begin{equation}
	\label{eq:LZ_complexity}
	\hat{H}(X) = \frac{n \log_2 n}{\sum_i \Lambda_i},
\end{equation}
where $n$ denotes the length of the time series, and $\Lambda_i$ denotes the 
length of the shortest substring starting from time $i$ 
that has not yet been observed prior to time $i$, i.e. from times $1$ to 
$t-1$. 
It is known that for stationary ergodic processes, $\hat{H}(X)$ converges to 
the entropy rate $H(X)$ almost surely \citep{Kontoyiannis1998} as 
$n \rightarrow \infty$. 

To estimate the joint entropy rate, one can extend \eqref{eq:LZ_complexity} 
to two time series, with $\Lambda_i$ denoting the length of the shortest 
substring of ordered pairs from both time series. 
This joint Lempel-Ziv complexity $\hat{H}(X,Y)$ also converges to the joint 
entropy rate as $n \rightarrow \infty$ \citep{Zozor2005}. 
I obtain an estimate of the 
\emph{conditional entropy rate} using the conditional Lempel-Ziv complexity 
$\hat{H}(X|Y) = \hat{H}(X,Y) - \hat{H}(Y)$. 

\section{Results}
I investigate the predictability of social interactions on two data sets. 
The first is the Reality Mining data \citep{Eagle2006}, which provides 
location (via nearest cellular tower) and interaction (via Bluetooth 
proximity) data for $94$ individuals at $5$-minute intervals over a year. 
The second is the Friends and Family data \citep{Aharony2011}, which 
provides only interaction (via Bluetooth proximity) data for $146$ individuals 
at $6$-minute intervals over $9$ months. 
Similar to \citet{Song2010}, 
I compare the estimated entropy rate $\hat{H}$ (using the Lempel-Ziv 
complexity) with 
the estimated entropy rates of an iid sequence with the same marginal  
probabilities as the observed sequence $\hat{H}_{iid}$ and an iid sequence of 
uniformly likely outcomes $\hat{H}_{unif}$.

\begin{figure}[t]
	\centering
	\includegraphics[width=2.2in]{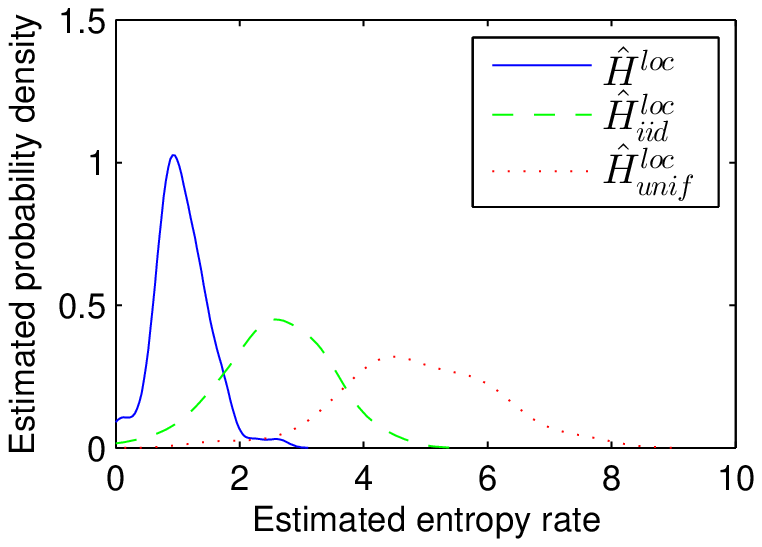}
	\qquad
	\includegraphics[width=2.2in]{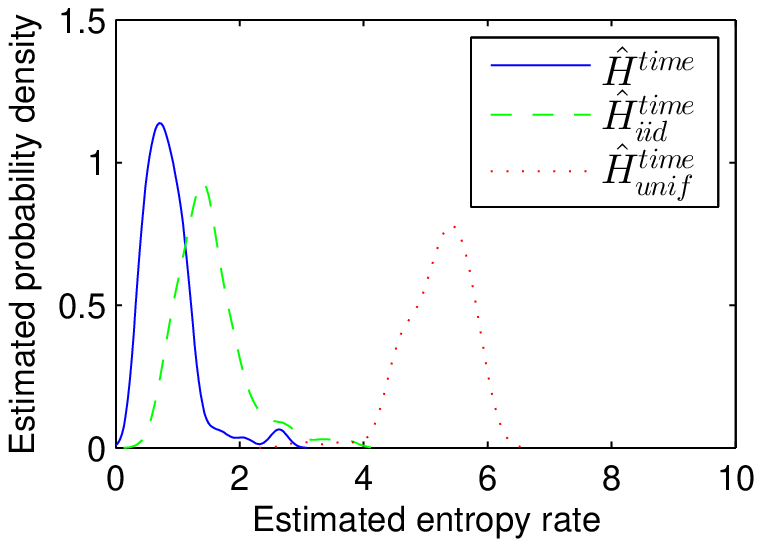}
	\caption{Distributions of the estimated
		entropy rate $\hat{H}$, iid entropy rate $\hat{H}_{iid}$, and 
		uniform entropy rate $\hat{H}_{unif}$ for locations (\emph{left}) 
		and times (\emph{right}) of interactions. 
	The low $\hat{H}^{loc}$ and $\hat{H}^{time}$ indicate that 
		\emph{locations and times of interactions are highly predictable}.}
	\label{fig:Entropy_locations}
\end{figure}

I begin with the Reality Mining data. 
The estimated entropy rates for the locations of interactions are shown in 
Fig.~\ref{fig:Entropy_locations} (left). 
The mean of $\hat{H}^{loc}$ is about $1.1$, indicating that the actual 
uncertainty in the location of an interaction is about $2^{1.1}=2.1$ 
locations. 
This is similar to the finding of \citet{Song2010} that 
the estimated entropy rate of an individual's location peaks at about $0.8$. 
Thus I conclude that the \emph{locations} of an individual's interactions are 
\emph{highly predictable}. 
The estimated entropy rate $\hat{H}^{loc}$ is much lower than 
the iid entropy rate $\hat{H}^{loc}_{iid}$, indicating that the temporal 
sequence is highly dependent. 
The estimated entropy rates for the times between interactions are shown in 
Fig.~\ref{fig:Entropy_locations} (right). 
Similar to locations, the \emph{times} of interactions are also 
\emph{highly predictable}. 

\begin{figure}[!t]
	\centering
	\includegraphics[width=2.2in]{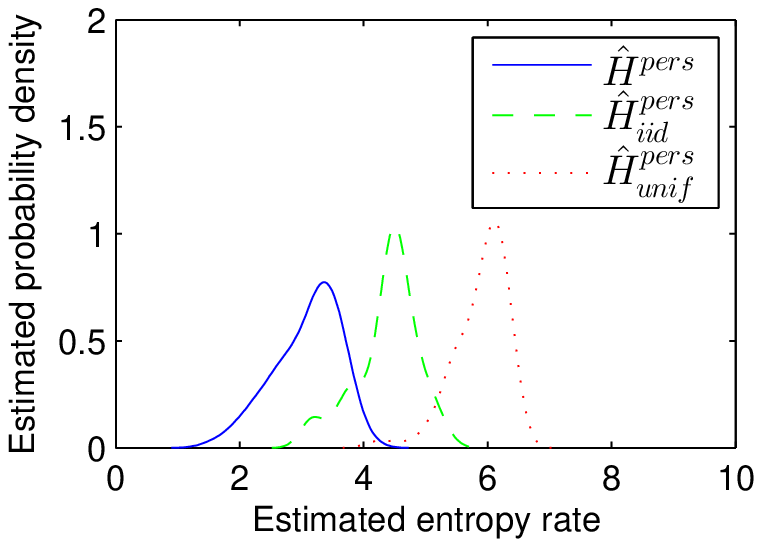}
	\qquad
	\includegraphics[width=2.2in]{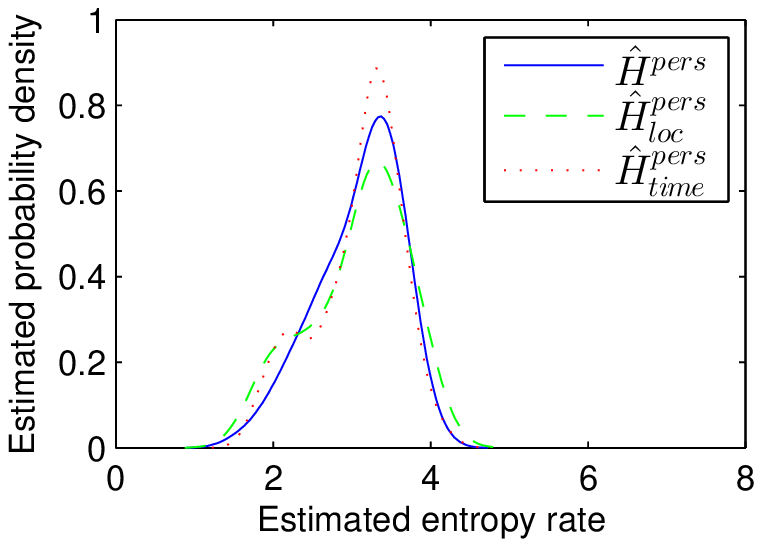}
	\caption{\emph{Left:} Distributions of the estimated
		entropy rate $\hat{H}^{pers}$, iid entropy rate 
		$\hat{H}^{pers}_{iid}$, and uniform entropy rate 
		$\hat{H}^{pers}_{unif}$. 
	$\hat{H}^{pers}$ is much higher than $\hat{H}^{loc}$ and 
		$\hat{H}^{time}$ (see Fig.~\ref{fig:Entropy_locations}), 
		indicating that the \emph{person an individual 
		interacts with is much less predictable} than location or time.
	\emph{Right:} Distributions of the estimated
		entropy rate $\hat{H}^{pers}$ and conditional entropy rates 
		given locations $\hat{H}^{pers}_{loc}$ and times 
		$\hat{H}^{pers}_{time}$. 
	There is little difference between the three distributions, 
		indicating that knowledge of times and locations \emph{does not 
		provide any significant benefit} in predicting the person an 
		individual interacts with.}
	\label{fig:Entropy_interactions}
\end{figure}

The estimated entropy rates for the person an individual interacts with are 
shown in Fig.~\ref{fig:Entropy_interactions} (left). 
Unlike with locations and times, the mean of $\hat{H}^{pers}$ is about 
about $3.1$, suggesting that the actual uncertainly is about $2^{3.1} = 8.5$ 
individuals. 
Thus it appears that the \emph{person} an individual interacts with is 
significantly 
\emph{less predictable} than the location or time! 
The estimated entropy rate $\hat{H}^{pers}$ is still much lower than the 
iid entropy rate $\hat{H}^{pers}_{iid}$, so some temporal dependency is 
still present in the time series.

Perhaps the person an individual interacts with may be more predictable if 
one is given the locations or times of interactions. 
The predictability given this additional information is captured by the 
\emph{conditional entropy rate}, which I estimate using the conditional 
Lempel-Ziv complexity $\hat{H}(X|Y) = \hat{H}(X,Y) - \hat{H}(Y)$. 
Somewhat surprisingly, I find that the estimated conditional entropy rates 
given locations $\hat{H}^{pers}_{loc}$ or times $\hat{H}^{pers}_{time}$ 
do not differ significantly from the estimated unconditional entropy 
rates, as shown in Fig.~\ref{fig:Entropy_interactions} 
(right)\footnote{Note that the true conditional entropy rate $H(X|Y)$ must 
always 
be less than the true unconditional entropy rate $H(X)$, but this is not 
necessarily true for the estimated conditional and unconditional entropy 
rates due to finite sample size.}. 
Thus I conclude that knowing the locations or times \emph{does not add much 
predictive 
value} when trying to predict the person with whom an individual interacts.

\begin{figure}[t]
	\centering
	\includegraphics[width=2.3in]{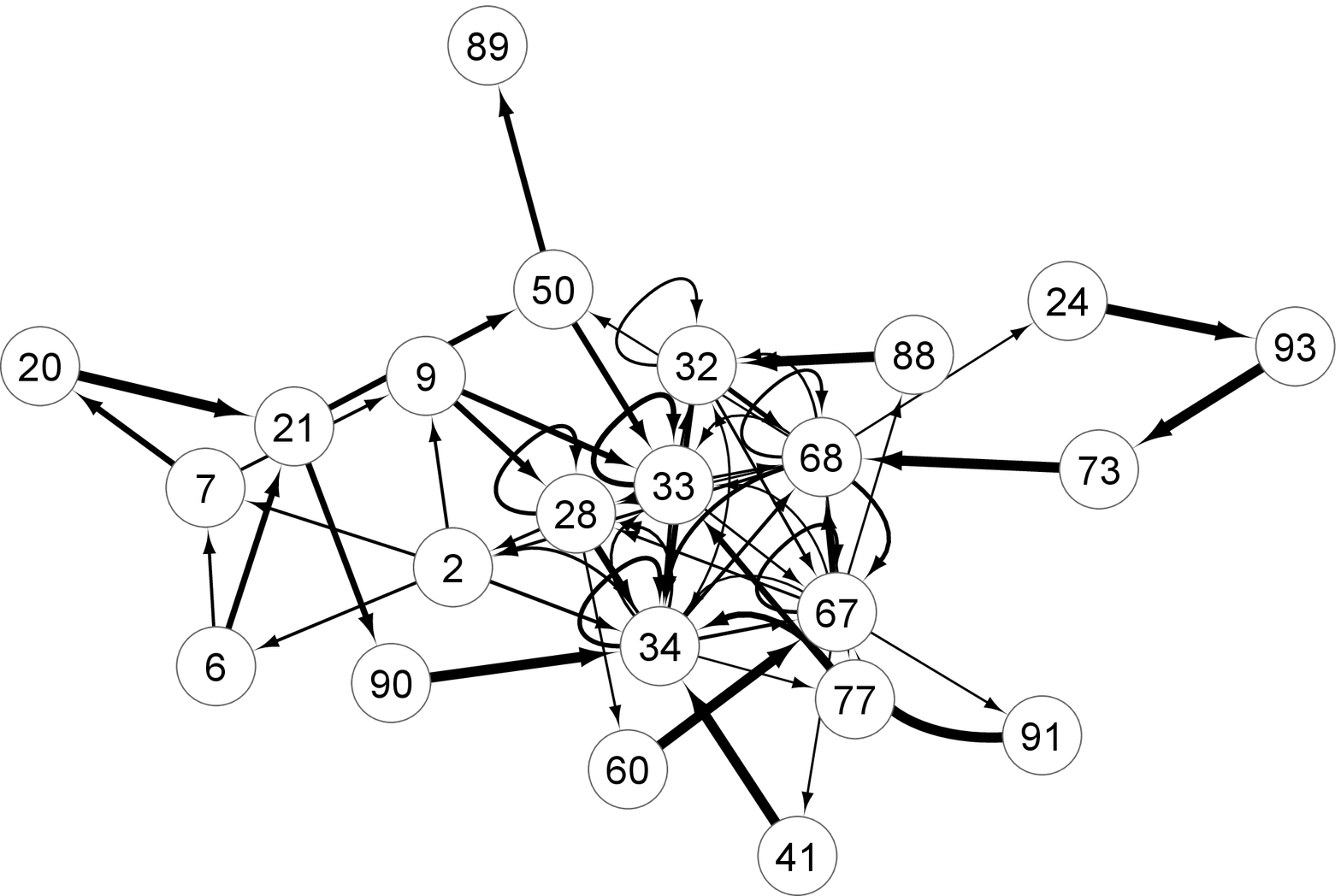}
	\qquad
	\includegraphics[width=2.2in]{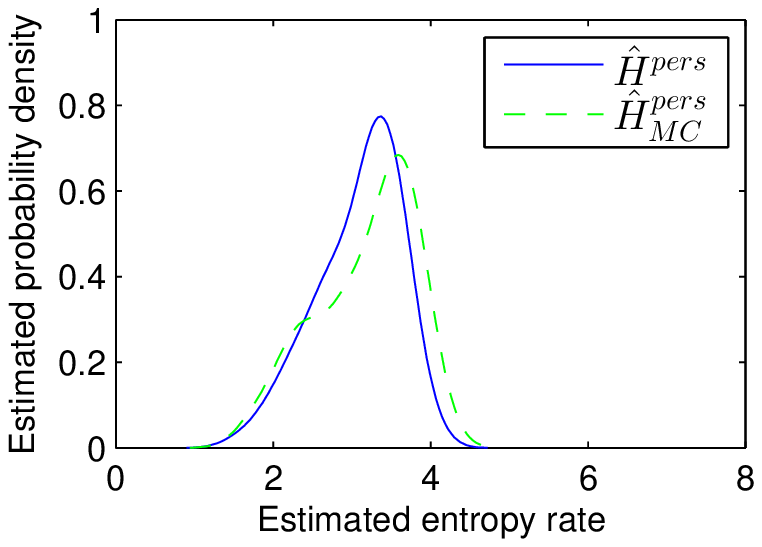}
	\caption{\emph{Left:} Graphical representation of Markov chain state 
		transition matrix for a selected individual. 
	Edge width is proportional to transition probability.
	\emph{Right:} Distributions of the estimated
		entropy rate $\hat{H}^{pers}$ and Markov chain entropy rate 
		$\hat{H}_{MC}^{pers}$. 
	The estimated entropy rates of the Markov chains are only slightly higher 
		than the rates of the actual sequences, suggesting that the Markov
		chain can achieve \emph{close to the upper bound} for predicting 
		the person an individual will interact with next.}
	\label{fig:Entropy_Markov}
\end{figure}

Entropy rates provide only an upper bound on predictability. 
I now consider the problem of actually modeling the sequence of people an 
individual interacts with over time. 
A simple model consists of a Markov chain, which assumes that the 
person an individual interacts with next depends only on the person she is 
currently interacting with (or the last person she interacted with, if she is 
not currently interacting with anyone). 
I learn a stationary Markov chain for each individual, such as the one 
pictured in Fig.~\ref{fig:Entropy_Markov} (left), and estimate the 
entropy rates of these chains by substituting relative frequencies for 
probabilities. 
As shown in Fig.~\ref{fig:Entropy_Markov} (right), the estimated 
entropy rates of the 
Markov chains $\hat{H}^{pers}_{MC}$ are only slightly 
higher\footnote{The true entropy 
rate is always lower than the Markov chain
entropy rate, but this is not necessarily true for the estimated 
rates, again, due to finite sample size.} 
than those of the actual sequences $\hat{H}^{pers}$. 
Specifically the mean $\hat{H}^{pers}_{MC}$ is $3.2$ compared to the mean 
$\hat{H}^{pers}$ of $3.1$. 
This suggests that the Markov chain is able to achieve \emph{close to the 
upper bound} for predicting the next person an individual interacts with! 

To measure how well the Markov chain works in practice, I learn 
the model on the first week of data and attempt to 
predict the most likely (top-$1$) and $5$ most likely (top-$5$) people an 
individual will interact with next for each interaction in the second week. 
I then update the model based on the interactions in the second week and 
repeat the process until I reach the end of the data. 
Overall, the predictions from the Markov chain achieved a top-$1$ accuracy 
of $19\%$ and top-$5$ accuracy of $49\%$. 
These results, while not spectacular, do appear to be reasonable given that 
I previously found the uncertainty to be about eight people.

I repeated the previous experiments on the Friends and Family data set. 
Location data is not available, but all of my findings from the Reality 
Mining data not involving location hold also for the Friends and Family data. 
The person an individual interacts with is slightly more predictable, 
with the mean $\hat{H}^{pers} = 2.3$ corresponding to uncertainty of 
about $2^{2.3} = 5.0$ people. 
The learned Markov chain model achieves mean $\hat{H}^{pers}_{MC} = 2.7$, 
which is again close to the mean $\hat{H}^{pers}$, although the larger gap 
compared to the Reality Mining data suggests that the effects of 
higher-order dependencies is stronger in the Friends and Family data. 
The predictions from the Markov chain achieved top-$1$ and top-$5$ 
accuracies of $21\%$ and $59\%$, respectively, which are also higher than 
in the Reality Mining data and agree with the lower entropy rate 
of the Friends and Family data. 

\section{Conclusions}
This study examined the predictability of social interactions, 
an important question in the emerging area of human dynamics. 
My main findings are threefold:
\begin{enumerate}
	\item The \emph{locations} and \emph{times} of interactions for an 
		individual are highly predictable, but not the \emph{other person} 
		with whom the individual interacts.
	\item Even if the locations and times of interactions are known,  
		there is \emph{almost no effect} on the predictability of the other 
		person.
	\item A simple Markov chain model achieves \emph{close to the upper 
		bound} for 
		predicting the next person with whom an individual will interact.
\end{enumerate}

I believe these findings have several key implications. 
Being able to predict the next person an individual will interact with could 
allow for indirect targeted marketing through this person. 
However, I found that there is significant uncertainty in who the 
next person is (roughly five to eight people), suggesting 
that one may need target a group of people rather than a single person. 
On a more positive note, the simplicity of the Markov chain model enables us 
to perform rigorous mathematical 
analyses that would not be possible with more complicated models. 
From the findings of this study, I believe that such a model is appropriate 
for making predictions about dynamic processes over social networks such as 
information diffusion and disease propagation.

\bibliographystyle{splncsnat}
\bibliography{library_sa}

\end{document}